\documentclass{aa}  

\usepackage{graphicx}
\usepackage{txfonts}
\usepackage{color}
\usepackage{stfloats}
\usepackage{amsmath}

\newcommand{\tablenotea}[1]{\parbox{9.0cm}{\indent \footnotesize{#1}}}
\newcommand{\tablenoteb}[1]{\parbox{18.3cm}{\indent \footnotesize{#1}}}

\newcommand{\ijp}{Indian J. Phys.}

\newcommand{\jctc}{J. Chem. Theory Comput.}
\newcommand{\jms}{J. Mol. Spectr.}
\newcommand{\jmst}{J. Mol. Struct.}

\newcommand{\jpca}{J. Phys. Chem. A}

\newcommand{\pccp}{PCCP}

\newcommand{\preva}{Phys. Rev. A}

\newcommand{\tfs}{Trans. Faraday Soc.}

\begin{document}

\title{First detection of doubly deuterated methyl acetylene\\ (CHD$_2$CCH and CH$_2$DCCD)\thanks{Based on observations carried out with the IRAM 30m Telescope. IRAM is supported by INSU/CNRS (France), MPG (Germany) and IGN (Spain).}}

\titlerunning{Doubly deuterated methyl acetylene}
\authorrunning{Ag\'undez et al.}

\author{M.~Ag\'undez\inst{1}, E.~Roueff\inst{2}, C. Cabezas\inst{1}, J.~Cernicharo\inst{1}, \and N.~Marcelino\inst{1}}

\institute{
Instituto de F\'isica Fundamental, CSIC, C/ Serrano 123, E-28006 Madrid, Spain\\
\email{marcelino.agundez@csic.es}
\and
LERMA, Observatoire de Paris, PSL Research University, CNRS, Sorbonne Universit\'e, F-92190, Meudon, France
}

\date{Received; accepted}

% \abstract{}{}{}{}{} 
% 5 {} token are mandatory
 
\abstract
% context heading (optional), leave it empty if necessary
{We report the first detection in space of the two doubly deuterated isotopologues of methyl acetylene. The species CHD$_2$CCH and CH$_2$DCCD were identified in the dense core L483 through nine and eight, respectively, rotational lines in the 72-116 GHz range using the IRAM 30m telescope. The astronomical frequencies observed here were combined with laboratory frequencies from the literature measured in the 29-47 GHz range to derive more accurate spectroscopic parameters for the two isotopologues. We derive beam-averaged column densities of (2.7\,$\pm$\,0.5)\,$\times$\,10$^{12}$ cm$^{-2}$ for CHD$_2$CCH and (2.2\,$\pm$\,0.4)\,$\times$\,10$^{12}$ cm$^{-2}$ for CH$_2$DCCD, which translate to abundance ratios CH$_3$CCH/CHD$_2$CCH\,=\,34\,$\pm$\,10 and CH$_3$CCH/CH$_2$DCCD\,=\,42\,$\pm$\,13. The doubly deuterated isotopologues of methyl acetylene are only a few times less abundant than the singly deuterated ones, concretely around 2.4 times less abundant than CH$_3$CCD. The abundances of the different deuterated isotopologues with respect to CH$_3$CCH are reasonably accounted for by a gas-phase chemical model in which deuteration occurs from the precursor ions C$_3$H$_6$D$^+$ and C$_3$H$_5$D$^+$, when the ortho-to-para ratio of molecular hydrogen is sufficiently low. This points to gas-phase chemical reactions, rather than grain-surface processes, as responsible for the formation and deuterium fractionation of CH$_3$CCH in L483. The abundance ratios CH$_2$DCCH/CH$_3$CCD\,=\,3.0\,$\pm$\,0.9 and CHD$_2$CCH/CH$_2$DCCD\,=\,1.25\,$\pm$\,0.37 observed in L483 are consistent with the statistically expected values of three and one, respectively, with the slight overabundance of CHD$_2$CCH compared to CH$_2$DCCD being well explained by the chemical model.}


\maketitle

\section{Introduction}

Deuterium fractionation is an extraordinary tool to study the early stages of star formation, in particular, the pre-stellar core phase, where the cold temperatures and high CO depletion offer optimal conditions to enrich molecules in deuterium, and the protostellar Class\,0 phase, where the gas displays some of the largest levels of deuteration observed as a consequence of the inheritance of gas and ice molecules enriched in deuterium during the pre-stellar core phase \citep{Ceccarelli2014}. The cosmic D/H ratio is $1.5\times10^{-5}$ \citep{Linsky2003}, but deuterium fractionation is so efficient in these environments that deuterated molecules can reach abundances as high as 30-40~\% relative to the parent hydrogenated species, as occurs for HDCS \citep{Marcelino2005} and CH$_2$DOH \citep{Parise2006}. Moreover, for abundant molecules containing more than one hydrogen atom, multiply deuterated forms have been observed with abundance enhancements of many orders of magnitude over the statistically expected value.

Molecules with two deuterium atoms observed to date comprise D$_2$CO \citep{Turner1990,Ceccarelli1998}, NHD$_2$ \citep{Roueff2000}, CHD$_2$OH \citep{Parise2002}, D$_2$S \citep{Vastel2003}, D$_2$H$^+$ \citep{Vastel2004,Parise2011}, D$_2$CS \citep{Marcelino2005}, D$_2$O \citep{Butner2007}, $c$-C$_3$D$_2$ \citep{Spezzano2013}, CHD$_2$CN \citep{Calcutt2018}, HCOOCHD$_2$ \citep{Manigand2019}, and ND$_2$ \citep{Melosso2020,Bacmann2020}. Even a couple of triply deuterated molecules have been detected: ND$_3$ \citep{Lis2002,vanderTak2002,Roueff2005} and CD$_3$OH \citep{Parise2004}. Multiply deuterated molecules have abundances relative to the hydrogenated species ranging from $5\times10^{-5}$, in the case of D$_2$O for hot corino conditions \citep{Butner2007}, to values of the order of 10~\% in cold dark cloud environments, which may become comparable to those of the corresponding singly deuterated species, as occurs for D$_2$CS \citep{Marcelino2005,Vastel2018,Agundez2019} and D$_2$CO \citep{Parise2006}. It is remarkable that in the case of the ion H$_3^+$, whose deuterated species drive most of the deuterium fractionation in the gas phase, the doubly deuterated form (first detected unambiguously by \citealt{Parise2011}) has an abundance similar to that of the singly deuterated species. Moreover, chemical models predict that the different deuterated species H$_2$D$^+$, D$_2$H$^+$, and D$_3^+$ become occasionally more abundant than H$_3^+$ itself for some time intervals and specific physical conditions \citep{Roberts2004,Sipila2017}.

Multiply deuterated molecules allow to put constraints on the deuteration mechanism, either in the gas phase or in the surface of dust grains, and by extension on the formation process of the precursor hydrogenated molecule. If deuterium fractionation occurs in the gas phase, the resulting abundance ratios between singly, doubly, and even triply deuterated forms of a given molecule vary depending on the abundances of H$_2$D$^+$, D$_2$H$^+$, and D$_3^+$ and the efficiency of the reactions of deuteron transfer from these ions, which in turn depend on parameters like the gas kinetic temperature, the level of CO depletion, and the ortho-to-para ratio of H$_2$ \citep{Roberts2004,Roueff2005,Flower2006}. If deuteration occurs on grains surfaces, the relative abundances of the different deuterated forms of a molecule are expected to follow a statistical pattern as they should basically depend on the relative arrival rates on grains of D and H atoms from the gas phase and thus on the abundance ratio between atomic D and H in the gas phase \citep{Brown1989}.

\begin{table}
\small
\caption{Observed line parameters of CHD$_2$CCH and CH$_2$DCCD in L483.}
\label{table:lines}
\centering
\begin{tabular}{crccc}
\hline \hline
\multicolumn{1}{c}{Transition} & \multicolumn{1}{c}{Frequency} & \multicolumn{1}{c}{$\Delta v$}      & \multicolumn{1}{c}{$T_A^*$ peak} & \multicolumn{1}{c}{$\int T_A^* dv$} \\
\multicolumn{1}{c}{}                & \multicolumn{1}{c}{(MHz)}        & \multicolumn{1}{c}{(km s$^{-1}$)}  & \multicolumn{1}{c}{(mK)}                   & \multicolumn{1}{c}{(mK km s$^{-1}$)} \\
\hline
\multicolumn{5}{c}{CHD$_2$CCH} \\
\hline
5$_{1,5}$-4$_{1,4}$      &  76643.616 &  0.47(6) & 12.9 & 6.5(6) \\ % rms = 1.5 mK
5$_{0,5}$-4$_{0,4}$      &  76979.250 &  0.37(4) & 33.0 & 13.1(11) \\ % rms = 2.6 mK
5$_{1,4}$-4$_{1,3}$      &  77317.279 &  0.42(5) & 21.5 & 9.7(11) \\ % rms = 2.5 mK
6$_{1,6}$-5$_{1,5}$      &  91971.307 &  0.29(4) & 17.4 & 5.3(8) \\ % rms = 2.5 mK
6$_{0,6}$-5$_{0,5}$      &  92372.787 &  0.27(2) & 31.0 & 8.9(7) \\ % rms = 2.4 mK
6$_{1,5}$-5$_{1,4}$      &  92779.672 &  0.30(6) & 17.4 & 5.6(9) \\ % rms = 2.1 mK
7$_{1,7}$-6$_{1,6}$      & 107298.386 &  0.30(6) & 13.7 & 4.4(7) \\ % rms = 2.1 mK
7$_{0,7}$-6$_{0,6}$      & 107765.046 &  0.29(3) & 28.3 & 8.7(9) \\ % rms = 2.9 mK
7$_{1,6}$-6$_{1,5}$      & 108241.467 &  0.39(7) & 15.5 & 6.5(9) \\ % rms = 2.8 mK
\hline
\multicolumn{5}{c}{CH$_2$DCCD} \\
\hline
5$_{1,5}$-4$_{1,4}$      &  73589.402 &  0.31(5) & 12.5 & 4.2(7) \\ % rms = 2.7 mK
5$_{0,5}$-4$_{0,4}$      &  73861.275 &  0.35(5) & 24.8 & 9.1(13) \\ % rms = 3.1 mK
5$_{1,4}$-4$_{1,3}$      &  74133.292 &  0.36(10) & 11.4 & 4.4(11) \\ % rms = 2.8 mK
6$_{1,6}$-5$_{1,5}$      &  88306.511 &  0.42(9) & 12.0 & 5.4(9) \\ % rms = 1.9 mK
6$_{1,5}$-5$_{1,4}$      &  88959.161 &  0.36(8) & 15.3 & 5.9(9) \\ % rms = 2.3 mK
7$_{1,7}$-6$_{1,6}$      & 103023.214 &  0.35(6) & 11.4 & 4.2(6) \\ % rms = 1.9 mK
7$_{0,7}$-6$_{0,6}$      & 103402.198 &  0.34(3) & 23.6 & 8.5(6) \\ % rms = 1.7 mK
7$_{1,6}$-6$_{1,5}$      & 103784.607 &  0.41(9) & 8.9 & 3.9(6) \\ % rms = 1.8 mK
\hline
\end{tabular}
\tablenotea{Numbers in parentheses are 1$\sigma$ uncertainties in units of the last digits. Line parameters were derived from a Gaussian fit to the line profile. Observed frequencies are obtained adopting $V_{\rm LSR}$ = 5.30 km s$^{-1}$ \citep{Agundez2019} and have an uncertainty of 20 kHz for all the lines. $\Delta v$ is the full width at half maximum.}
\end{table}

For molecules with various hydrogen atoms among which some of them are equivalent, the relative abundances of the different deuterated variants can also give clues on the formation of the molecule. For example, for methanol (CH$_3$OH) deuteration on the methyl group is three times more probable than on the hydroxyl group, but observed CH$_2$DOH/CH$_3$OD ratios differ significantly from the statistical value of three, with values well above or below depending on the source \citep{Parise2006,Ratajczak2011,Agundez2019}. The case of methyl acetylene (CH$_3$CCH) is also interesting as, similarly to methanol, it contains a methyl group with three equivalent H atoms and one non-equivalent H atom in the terminal CCH group. The two singly deuterated forms of CH$_3$CCH, CH$_2$DCCH and CH$_3$CCD, were first detected in the interstellar medium by \cite{Gerin1992} and \cite{Markwick2005}, respectively. Using the ARO 12m telescope, \cite{Markwick2005} reported CH$_2$DCCH/CH$_3$CCD ratios in the range 1.20-2.04 along the TMC-1 ridge, significantly different from the statistical value of three, although in L483 the observed ratio is consistent with the statistical value \citep{Agundez2019}.

Here we report the first identification in space of the two forms of doubly deuterated methyl acetylene: CHD$_2$CCH and CH$_2$DCCD. These species were detected toward the dense core L483 using the IRAM 30m telescope. The observation of the two singly and two doubly deuterated forms of methyl acetylene stimulated us to build a gas-phase chemical model, including multiply deuterated species, to investigate the mechanisms of deuteration of CH$_3$CCH in L483.

\section{Observations}

The target of the observations, L483, is a dark cloud core located in the Aquila Rift star-forming region, which hosts an embedded infrared source, IRAS\,18148$-$0440, classified as a Class\,0 object. The observations were carried out with the IRAM 30m telescope in the frame of a $\lambda$~3 mm line survey covering the 80-116 GHz range \citep{Agundez2019}. Additional observations in the 72-80 GHz range were taken with the IRAM 30m telescope in December 2018. During this observing run, weather conditions were good, with 4-5 mm of precipitable water vapor.

We describe briefly the observations. For more details we refer to \cite{Agundez2019}. We used the EMIR receiver E090 connected to a fast Fourier transform spectrometer providing a spectral resolution of 50 kHz. All observations were carried out using the frequency-switching technique, with a frequency throw of 7.2 MHz. Intensities are expressed in terms of $T_A^*$, the antenna temperature corrected for atmospheric absorption and for antenna ohmic and spillover losses. The uncertainty in $T_A^*$ is estimated to be 10~\%. The antenna temperature $T_A^*$ can be converted to main beam brightness temperature $T_{\rm mb}$ by dividing by $B_{\rm eff}$/$F_{\rm eff}$, where $B_{\rm eff}$ = 0.871 $\exp{[-(\nu{\rm (GHz)}/359)^2]}$ and $F_{\rm eff}$ = 0.95\footnote{\texttt{http://www.iram.es/IRAMES/mainWiki/Iram30mEfficiencies}}. Pointing errors were typically 2-3$''$ and the half power beam width of the IRAM 30m telescope ranges between 34$''$ at 72 GHz and 21$''$ at 116 GHz. All data were reduced using the program {\small CLASS} of the {\small GILDAS} software package\footnote{\texttt{http://www.iram.fr/IRAMFR/GILDAS}}.

\section{Results}

\subsection{Detection of CHD$_2$CCH and CH$_2$DCCD in L483}

\begin{figure*}
\centering
\includegraphics[angle=0,width=0.90\textwidth]{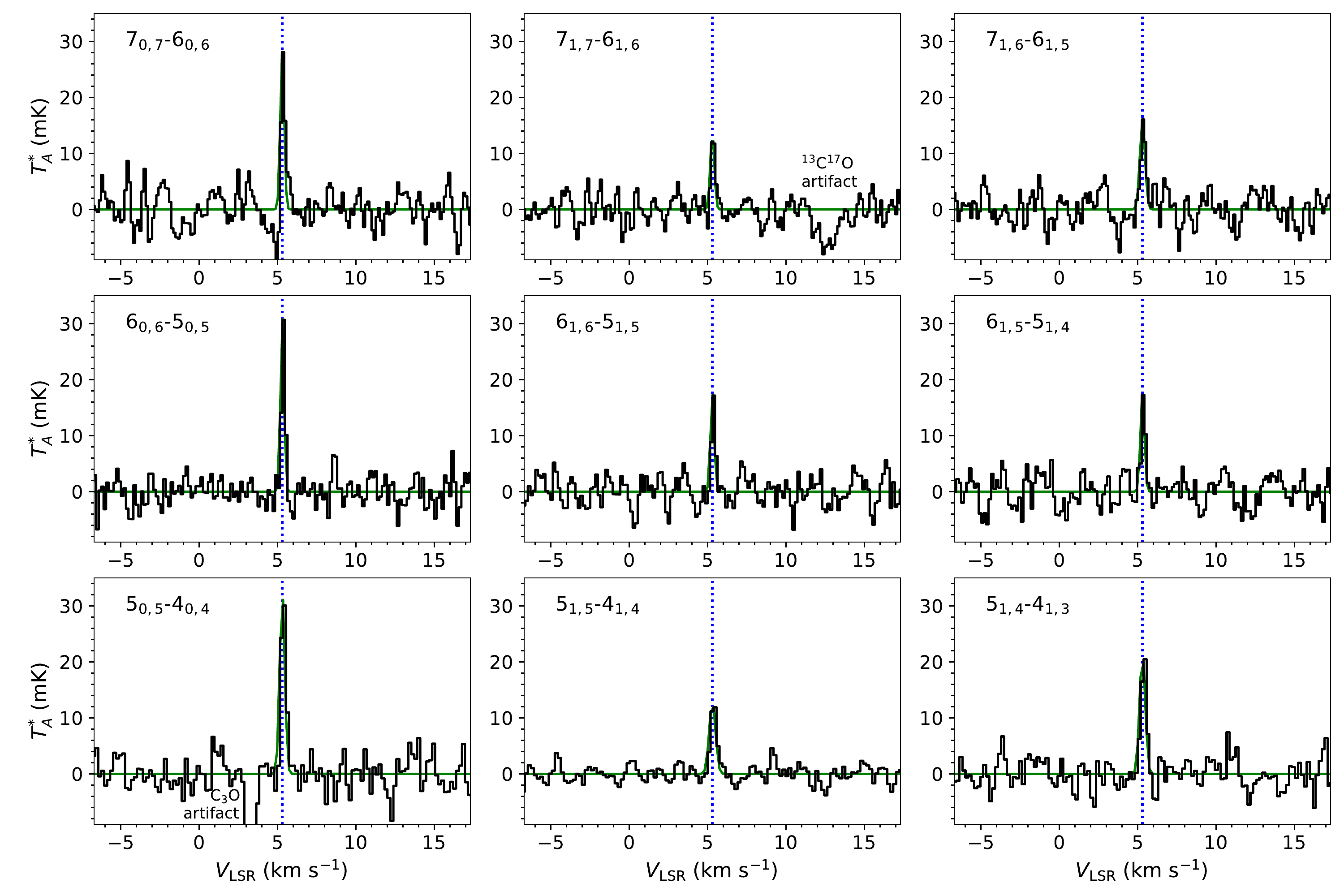}
\caption{Observed lines of CHD$_2$CCH in L483 in the 72-116 GHz range. The systemic velocity of L483 (5.30 km s$^{-1}$) is indicated by a vertical dotted blue line. Green lines are Gaussian fits to each line profile. Frequencies and line parameters are given in Table~\ref{table:lines}.} \label{fig:lines_chd2cch}
\end{figure*}

\begin{figure*}
\centering
\includegraphics[angle=0,width=0.90\textwidth]{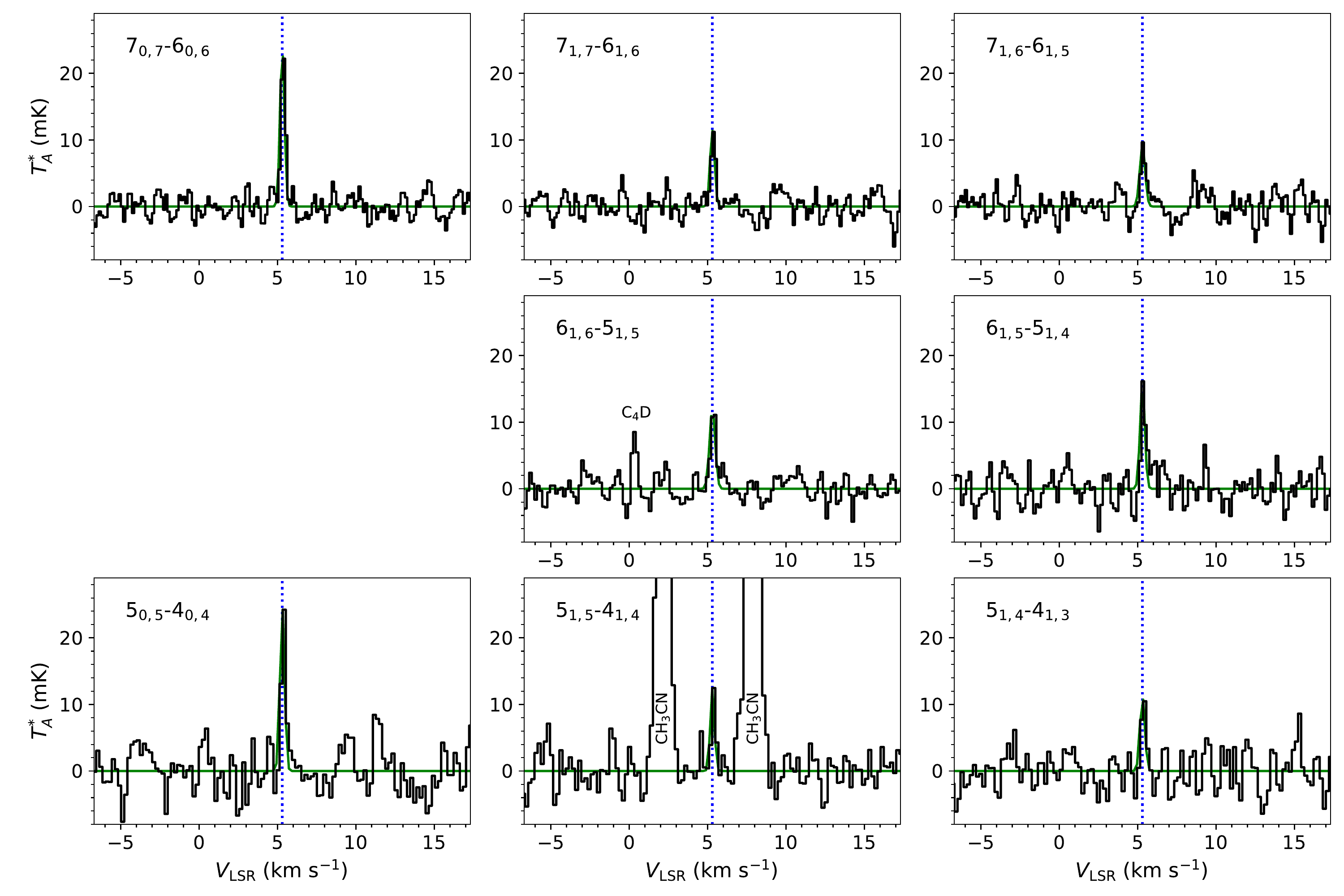}
\caption{Observed lines of CH$_2$DCCD in L483 in the 72-116 GHz range. The systemic velocity of L483 (5.30 km s$^{-1}$) is indicated by a vertical dotted blue line. Green lines are Gaussian fits to each line profile. Frequencies and line parameters are given in Table~\ref{table:lines}. The $6_{0,6}$-$5_{0,5}$ line is not detected because it overlaps with the strong HCN $J=1$-0 line.} \label{fig:lines_ch2dccd}
\end{figure*}

In the data of L483 covering the 72-116 GHz frequency range there are several lines which could not be assigned to any known molecule after inspection of the 
{\small CDMS}\footnote{\texttt{https://cdms.astro.uni-koeln.de/}} \citep{Muller2005} and {\small JPL}\footnote{\texttt{https://spec.jpl.nasa.gov/}} \citep{Pickett1998} catalogues. Several of these unidentified lines could be grouped in sets of three lines with harmonic relations 5/6/7, with rotational constants around either 7700 MHz or 7400 MHz. These values are close to the rotational constants $B$ or $(B+C)/2$ of the isotopologues of methyl acetylene. For example, for CH$_3$CCD the rotational constant $B$ is 7788 MHz and for CH$_2$DCCH the value of $(B+C)/2$ is 8091 MHz \citep{LeGuennec1993}. A closer inspection of the private catalogue of J. Cernicharo generated from the {\small MADEX} code\footnote{\texttt{https://nanocosmos.iff.csic.es/?page\_id=1619}} \citep{Cernicharo2012} revealed that the observed frequencies of these sets of harmonically related lines differ by 0.9-2.7 MHz and 0.2-1.7 MHz from the frequencies predicted for rotational transitions of CHD$_2$CCH and CH$_2$DCCD, respectively. The rotational spectrum of these two molecules has been measured in the laboratory in the frequency range 29-47 GHz with uncertainties of 0.1 MHz in the frequencies \citep{Thomas1955}. Using these data, the frequencies predicted for the 72-116 GHz range have errors of a few MHz, and this is the reason why some of the unidentified lines seen in the 80-116 GHz range were not properly assigned to CHD$_2$CCH and CH$_2$DCCD by \cite{Agundez2019}. In fact, the identification of these two molecules reduces drastically the number of lines pending of identification in the $\lambda$ 3 mm line survey of L483 (see Table A.2 of \citealt{Agundez2019}). The parameters derived for the lines assigned to the two doubly deuterated isotopologues of methyl acetylene are given in Table~\ref{table:lines} and the observed line profiles are shown in Fig.~\ref{fig:lines_chd2cch} for CHD$_2$CCH and in Fig.~\ref{fig:lines_ch2dccd} for CH$_2$DCCD. We note that the line at 91971.307 MHz, assigned to the 6$_{1,6}$-5$_{1,5}$ transition of CHD$_2$CCH, was incorrectly assigned to the transition 5$_3$-4$_3$ $F$\,=\,6-5 of CH$_3$CN by \cite{Agundez2019}.

The astronomical frequencies observed in the 72-116 GHz range (see Table~\ref{table:lines}) and the laboratory frequencies measured in the 29-47\,GHz range \citep{Thomas1955} were used to derive more accurate spectroscopic parameters for CHD$_2$CCH and CH$_2$DCCD (see Table~\ref{table:constants}). These constants improve considerable over those reported by \cite{Thomas1955} and allow to predict the rotational spectrum of the two molecules with an accuracy better than 0.1 MHz up to 200 GHz. Since the lines of doubly deuterated CH$_3$CCH are relatively bright in L483, it would not be surprising to detect the triply deuterated species CHD$_2$CCD, which should be three times more abundant than CD$_3$CCH. However, the rotational spectrum has only been measured in the 28-43\,GHz range with a modest accuracy \citep{Thomas1955}, which may translate to frequency errors of a few MHz for lines lying above 72\,GHz. It would therefore be interesting to measure the rotational spectrum of triply deuterated CH$_3$CCH at millimeter wavelengths.

\begin{table}
\small
\caption{Spectroscopic parameters of CHD$_2$CCH and CH$_2$DCCD.}
\label{table:constants}
\centering
\begin{tabular}{lcc}
\hline \hline
\multicolumn{1}{c}{Parameter} & \multicolumn{1}{c}{CHD$_2$CCH} & \multicolumn{1}{c}{CH$_2$DCCD} \\
\hline
$A_0$ (MHz) ......................... & 95746(533) & 119688.0(170) \\        
$B_0$ (MHz) ......................... & 7765.71927(255) & 7440.77759(341) \\        
$C_0$ (MHz) ......................... & 7630.99013(254) & 7332.00345(340) \\        
$\Delta_J$ (kHz) .......................... & 2.3807(270) & 2.0759(347) \\        
$\Delta_{JK}$ (kHz) ........................ & 123.83(224) & 118.93(295) \\        
\hline
$N$\,$^a$ ................................... & 16 & 17 \\        
rms (kHz)\,$^b$ ...................... & 52 & 80 \\        
\hline
\end{tabular}
\tablenotea{Numbers in parentheses are 1$\sigma$ uncertainties in units of the last digits.\\
$^a$\,Number of lines included in the fit. For CHD$_2$CCH, $J_{\rm max}$\,=\,7,  $K_{\rm max}$\,=\,2, and $\nu_{\rm max}$\,=\,108.241\,GHz, and for CH$_2$DCCD, $J_{\rm max}$\,=\,7,  $K_{\rm max}$\,=\,2, and $\nu_{\rm max}$\,=\,103.784\,GHz. $^b$\,Root mean square of the fit.}
\end{table}

%Para CHD2CCH NLines=16 sigma=52 kHz Jmax=7 Kmax=2 vmax=108.241 GHz
%Para CH2DCCD Nlines=17 sigma=80 kHz Jmax=7 Kmax=2 vmax=103.784 GHz

\subsection{Observed column densities}

We derived beam-averaged column densities for methyl acetylene and its singly and doubly deuterated isotopologues in L483 assuming local thermodynamic equilibrium. Values for CH$_3$CCH and the two singly deuterated forms were already reported by \cite{Agundez2019}. Here we have revised some of these values including new lines observed in the 72-80 GHz range and using new values for the electric dipole moments. 

\begin{table*}[hb!]
\caption{Column densities and ratios for CH$_3$CCH and its singly and doubly deuterated species in L483.} \label{table:column_densities}
\small
\centering
\begin{tabular}{lrrrcrc@{\hspace{2.0cm}}lc}
\hline \hline
\multicolumn{1}{l}{Species} & \multicolumn{1}{c}{$\mu$ (D)\,$^a$} & \multicolumn{1}{c}{$N_{\rm lines}$\,$^e$} & \multicolumn{1}{c}{$E_{\rm up}$ (K)\,$^f$} & \multicolumn{1}{c}{$T_{\rm rot}$ (K)} & \multicolumn{1}{r}{$N$ (cm$^{-2}$)\,$^h$} & & \multicolumn{2}{c}{Column density ratio\,$^i$} \\
\hline
% CH3CCH(K=0) = 4.411080e13 
% CH3CCH(K=1) = 3.987523e13
% CH3CCH(K=2) = 6.030911e12
% CH3CCH(K=3) = 9.646669e11 
% CH3CCH(A, K>3 for T=15 K) = 1.20e6
% CH3CCH(E, K>2 for T=15 K) = 1.84e10
% N( all K) = N(K=0,1) x 1.083514
CH$_3$CCH     & 0.7839\,$^b$ & 8 & 11.5 - 82.3 & 10.2\,$^g$ & $9.10\times10^{13}$       & & & \\ % 9.100001e13
CH$_2$DCCH  & 0.7850\,$^c$ & 15 & 11.6 - 43.6 & $10.2 \pm 0.7$  & $(1.76 \pm 0.34)\times10^{13}$ & & CH$_3$CCH / CH$_2$DCCH & $5.2 \pm 1.6$ \\ % 1.756114e13
% CH3CCD(K=0) = 2.790177e12
% CH3CCD(K=1) = 2.555638e12
CH$_3$CCD    & 0.7698\,$^d$ & 6 & 10.5 - 20.9 & 10.2\,$^g$  & $5.79\times10^{12}$      & & CH$_3$CCH / CH$_3$CCD    & $15.7 \pm 4.7$ \\ % 5.792265e12 % A/E = 1.09
CHD$_2$CCH & 0.7870\,$^c$ & 9 & 11.1 - 25.0 & 10.2\,$^g$ & $2.69\times10^{12}$        & & CH$_3$CCH / CHD$_2$CCH & $33.8 \pm 10.1$ \\ % 2.685539e12 (Trot = 9.28 +/- 1.31 K; consistent with 10.2 K)
CH$_2$DCCD & 0.7711\,$^c$ & 8 & 10.6 - 25.3 & 10.2\,$^g$ & $2.16\times10^{12}$        & & CH$_3$CCH / CH$_2$DCCD & $42.1 \pm 12.6$ \\ % 2.158584e12 (Trot = 11.39 +/- 1.99 K; consistent with 10.2 K)
\hline
\end{tabular}
\tablenoteb{$^a$\,Dipole moments along the principal axis $a$. $^b$\,Experimental value from \cite{Burrell1980}. $^c$\,Calculated value (this work). $^d$\,Experimental value from \cite{Muenter1966} corrected for a more accurate dipole moment of the reference OCS than used by these authors (see text). $^e$\,Number of lines observed. $^f$\,Range of upper level energies of the lines observed. $^g$\,Rotational temperature has been fixed to the value derived for CH$_2$DCCH. $^h$\,Errors in the column densities are estimated to be 20~\%. $^i$\,Errors in the column density ratios are estimated to be 30~\%.}
\end{table*}

The dipole moment of CH$_3$CCH was experimentally measured by \cite{Burrell1980} to be 0.7839 D. For CH$_3$CCD, the dipole moment was measured by \cite{Muenter1966}. Here we adopt the value measured by these latter authors corrected by using a more accurate value of the dipole moment of OCS, which is used as reference (see \citealt{Burrell1980}). The dipole moment of CH$_3$CCD turns out to be only $\sim2~\%$ lower than that of CH$_3$CCH (see Table~\ref{table:column_densities}). For CH$_2$DCCH, CHD$_2$CCH, and CH$_2$DCCD, all of which are asymmetric rotors, dipole moments are not known, and we thus evaluated them using quantum chemical calculations. Since for these species the dipole moment is only due to vibrational effects connected to isotopic substitution, anharmonic force field calculations were required. All the calculations were done using the orbital-dependent functional (dubbed B2PLYP) \citep{Grimme2006} with the Becke-Johnson D3(BJ) damping function \citep{Risthaus2013} and the Dunning's basis set cc-pVTZ \citep{Dunning1989}. In order to obtain more reliable values we also calculated the dipole moments for CH$_3$CCH, CD$_3$CCH, CH$_3$CCD and CD$_3$CCD, whose dipole moments are experimentally known \citep{Muenter1966,Burrell1980}. In this manner, we obtained an experimental/theoretical ratio that was used to correct the theoretical values for CH$_2$DCCH, CHD$_2$CCH, and CH$_2$DCCD. The calculated dipole moments along the principal axis $a$, which is the relevant parameter since the observed lines are of $a$-type, are given in Table~\ref{table:column_densities}. It is found that they differ only slightly, up to $\sim2~\%$, from that of CH$_3$CCH.

For CH$_3$CCD we now include two new lines, the 5$_0$-4$_0$ and 5$_1$-4$_1$, observed in the 72-80 GHz range. For CH$_3$CCH, we observed lines with $K=0, 1, 2, 3$, and thus the column density includes these four $K$ ladders, although the $K=0$ and $K=1$ levels account for most of the column density (92~\%). For CH$_3$CCD only $K=0$ and $K=1$ lines were observed. Among all isotopologues of methyl acetylene, CH$_2$DCCH has the most precise determination of the rotational temperature, 10.2\,$\pm$\,0.7 K, due to the higher number of lines observed and the wider range of upper level energies involved. We therefore adopted this value for all other isotopologues when deriving column densities. We note that if the rotational temperature is not fixed, the values obtained are consistent with the rotational temperature of 10.2 K adopted. For example, for CHD$_2$CCH we obtain a rotational temperature of 9.3 $\pm$ 1.3 K while for CH$_2$DCCD we get 11.4\,$\pm$\,2.0 K. The rotational partition function was computed by summation including energy levels up to $J$\,=\,100. For guidance, in Table~\ref{table:qrot} we give rotational partition functions for CHD$_2$CCH and CH$_2$DCCD at different temperatures. The column densities derived in L483 for CH$_3$CCH and its singly and doubly deuterated forms are given in Table~\ref{table:column_densities}. Deuterium ratios for various molecules in L483 have been discussed by \cite{Agundez2019}. Regarding doubly deuterated methyl acetylene, the two forms show a deuterium enrichment comparable to H$_2$CO and H$_2$CS, which show the largest levels of deuteration in L483 (H$_2$CO/D$_2$CO\,=\,31.3 and H$_2$CS/D$_2$CS\,=\,21.7; \citealt{Agundez2019}), and are significantly more enriched in deuterium than CH$_3$OH and $c$-C$_3$H$_2$, for which CH$_3$OH/CHD$_2$OH\,=\,357 and $c$-C$_3$H$_2$/$c$-C$_3$D$_2$\,=\,103 \citep{Agundez2019}.

\begin{table}
\small
\caption{Rotational partition functions for CHD$_2$CCH and CH$_2$DCCD.}
\label{table:qrot}
\centering
\begin{tabular}{c@{\hspace{1.7cm}}r@{\hspace{1.6cm}}r}
\hline \hline
Temperature (K) & \multicolumn{2}{c}{$Q_{rot}$\,$^a$} \\
\cline{2-3}
                                                          & CHD$_2$CCH & CH$_2$DCCD \\
\hline
300.0 & 11646.9629 & 10855.0235 \\        
225.0 & 7563.8563 & 7049.8187 \\        
150.0 & 4117.1979 & 3837.5133 \\        
75.00 & 1456.4144 & 1357.4949 \\        
37.50 & 515.6773 & 480.6346 \\        
18.75 & 182.8944 & 170.4479 \\        
9.375 & 65.0786 & 60.6361 \\        
5.000 & 25.6362 & 23.8844 \\        
2.725 & 10.5863 & 9.9782 \\        
\hline
\end{tabular}
\tablenotea{$^a$\,Computed including energy levels up to $J$\,=\,100.}
\end{table}

\section{Chemical model} \label{sec:model}

To our knowledge, detailed studies of multiple deuteration have been essentially focused on simple hydrides, like H$_2$O, NH$_2$, and NH$_3$ \citep{Sipila2015,Furuya2015,Roueff2005,Hily-Blant2018,Bacmann2020}, and molecules containing a single carbon atom, like H$_2$CO, H$_2$CS, and CH$_3$OH \citep{Marcelino2005,Parise2006,Taquet2012}. Multiple deuteration proceeds essentially in the gas phase through successive reactions of HD with H$_3^+$, H$_2$D$^+$, and D$_2$H$^+$, and with CH$_3^+$, CH$_2$D$^+$, and CHD$_2^+$, as introduced by \cite{Roberts2004}. Additional deuterium exchange reactions with HD suggested to be considered for carbon chain molecules involve C$_2$H$_2^+$ and C$_2$HD$^+$ \citep{Turner2001}. We further introduce exchange reactions of C$_3$H$^+$ with HD and D$_2$, which have been studied by \cite{SavicGerlich2005,Savic2005} as displayed in Table~\ref{table:reac_c3hplus}. We estimate the endothermicity for the reverse reactions from the values reported for the cyclic and the linear isomers of C$_3$H and C$_3$D \citep{Etim2020}.

We consider here a pure gas-phase chemical network including species containing up to three deuterium atoms, extending our previous study on the recent detection of HDCCN in \mbox{TMC-1} \citep{Cabezas2021}. As an example, let us mention that, in order to comply with the present observational challenges, we introduced the various isomers of deuterated methyl acetylene CH$_2$DCCH, CH$_3$CCD, CHD$_2$CCH, CH$_2$DCCD, CD$_3$CCH, and CHD$_2$CCD. The final chemical network comprises 394 species linked through more than 14,000 reactions. When no experimental information is available, we assume that the total rate coefficient of reactions involving deuterium atoms is identical to that with hydrogenated compounds. The derivation of branching ratios is principally based on statistical considerations.

\begin{table}
\footnotesize
\caption{Reactions  involving  C$_3$H$^+$ and C$_3$D$^+$.}
\label{table:reac_c3hplus}
\centering
\begin{tabular}{lllc}
\hline \hline
\multicolumn{1}{l}{Reaction} & \multicolumn{1}{c}{$\alpha$\,(cm$^{3}$s$^{-1}$)}   & \multicolumn{1}{c}{$\gamma$\,(K)} & \multicolumn{1}{c}{Ref.} \\
\hline
C$_3$H$^+$ + HD $\rightleftarrows$ C$_3$D$^+$ + H$_2$  & $5.6 \times 10^{-11}$ & 500 - 375 & (1) \\
C$_3$H$^+$ + D$_2$ $\rightleftarrows$ C$_3$D$^+$ + HD  & $3.0 \times 10^{-13}$ & 420 - 320 & (1) \\
\hline
C$_3^+$ + H$_2$ $\rightarrow$ C$_3$H$^+$ + H         & $1.7 \times 10^{-9}$  & & (1) \\
C$_3^+$ + HD $\rightarrow$ C$_3$D$^+$ + H            & $9.3 \times 10^{-10}$ & & (1) \\
C$_3^+$ + HD $\rightarrow$ C$_3$H$^+$ + D            & $7.6 \times 10^{-10}$ & & (1) \\
C$_3^+$ + D$_2$ $\rightarrow$ C$_3$D$^+$ + D         & $1.3 \times 10^{-9}$  & & (1) \\
C$_3$H$^+$ + H$_2$  $\rightarrow$ c-C$_3$H$_2^+$ + H & $1.0 \times 10^{-12}$ & & (2) \\
C$_3$H$^+$ + HD  $\rightarrow$ c-C$_3$HD$^+$ + H     & $4.6 \times 10^{-10}$ & & (1) \\
C$_3$H$^+$ + HD  $\rightarrow$ c-C$_3$H$_2^+$ + D    & $3.0 \times 10^{-12}$ & & (1) \\
C$_3$H$^+$ + D$_2$  $\rightarrow$ c-C$_3$HD$^+$ + D  & $1.0 \times 10^{-11}$ & & (1) \\
C$_3$H$^+$ + D$_2$  $\rightarrow$ c-C$_3$D$_2^+$ + H & $2.7 \times 10^{-11}$ & & (1) \\
C$_3$D$^+$ + H$_2$  $\rightarrow$ c-C$_3$HD$^+$ + H  & $1.0 \times 10^{-12}$ & & (3) \\
C$_3$D$^+$ + HD  $\rightarrow$ c-C$_3$HD$^+$ + H     & $1.0 \times 10^{-10}$ & & (1) \\
C$_3$D$^+$ + HD  $\rightarrow$ c-C$_3$D$_2^+$ + H    & $8.3 \times 10^{-11}$ & & (1) \\
C$_3$D$^+$ + D$_2$  $\rightarrow$ c-C$_3$D$_2^+$ + D & $1.7 \times 10^{-10}$ & & (1) \\
\hline
\end{tabular}
\tablenotea{The rate coefficient, $\alpha$, corresponds to the value reported at 10\,K or 15\,K  in reference (1). The first two displayed reactions correspond to deuterium exchange and $\gamma$ stands for the estimated endothermicity (see text) of the reverse reaction.\\
References: (1)\,\cite{SavicGerlich2005}; (2)\,\cite{Loison2017}; (3)\,assumed.
}
\end{table}

\begin{table}
\footnotesize
\caption{Reactions involving  ortho H$_2$.}
\label{table:reac_oh2}
\centering
%\begin{tabular}{|lll|c|c|r|}
\begin{tabular}{l@{\hspace{0.1cm}}rrrc}
\hline \hline
%\multicolumn{3}{|c|}{Reaction}   &  $\alpha$  (cm$^{3}$s$^{-1}$) &  $\beta$& $\gamma$ (K) \\
\multicolumn{1}{l}{Reaction} & \multicolumn{1}{c}{$\alpha$\,(cm$^{3}$s$^{-1}$)} & \multicolumn{1}{c}{$\beta$} & \multicolumn{1}{c}{$\gamma$\,(K)} & \multicolumn{1}{c}{\hspace{-0.2cm} Ref.} \\
\hline
$o$-H$_2$ + H$_2$D$^+$ $\rightarrow$ HD + H$_3^+$              &  $9.4 \times 10^{-11}$   &   $-$0.79     &   56.0 & (1)     \\
$o$-H$_2$ + D$_2$H$^+$ $\rightarrow$ HD + H$_2$D$^+$       &  $3.3 \times 10^{-10}$  &   $-$0.55  &    23.1 & (1)  \\
$o$-H$_2$ + D$_2$H$^+$ $\rightarrow$ D$_2$ + H$_3^+$        & $3.5 \times 10^{-11}$   & $-$0.82  &  181.7 & (1)   \\
$o$-H$_2$ + D$_3^+$        $\rightarrow$ HD + D$_2$H$^+$       &   $9.2 \times 10^{-10}$   & $-$0.59   &  68.8 & (1)  \\
$o$-H$_2$ + D$_3^+$        $\rightarrow$ D$_2$ + H$_2$D$^+$ &   $1.5 \times 10^{-10}$   & $-$0.85   &  174.4 & (1)   \\
$o$-H$_2$ + N$^+$            $\rightarrow$ NH$^+$ + H                 &   $4.2 \times 10^{-10}$   & $-$0.15   &  44.1 & (2)  \\
\hline
\end{tabular}
\tablenotea{Reaction rate coefficient is given by \mbox{$k(T) = \alpha\,(\frac{T}{300\,K})^{\beta}\,\exp{(-\gamma/T)}$}.\\
References: (1)\,\cite{Hily-Blant2018}; (2)\,\cite{Dislaire2012}.
}
\end{table}

However, for deuterated methyl acetylene we consider that the external CH and CCH (or CD and CCD) groups are much more mobile than the methyl group. Then, we assume that for reactions involving the ejection of CH and CCH (or CD and CCD), only the terminal group is implied in the reaction. As an example:
\begin{subequations} \label{reac:c++ch2dcch}
\begin{align}
\rm C^+ + CH_2DCCH & \rightarrow \rm C_3H_2D^+ + CH \\
\rm C^+ + CH_2DCCH & \rightarrow \rm C_2H_2D^+ + C_2H
\end{align}
\end{subequations}
\begin{subequations} \label{reac:c++ch3ccd}
\begin{align}
\rm C^+ + CH_3CCD & \rightarrow \rm C_3H_3^+ + CD \\
\rm C^+ + CH_3CCD & \rightarrow \rm C_3H_3^+ + C_2D
\end{align}
\end{subequations}
For reactions of dissociative recombination, we consider that atomic hydrogen is ejected twice more efficiently than deuterium, as found in some experiments (e.g., \citealt{Jensen1999}). We include the findings of \cite{SavicGerlich2005} and \cite{Savic2005}, who showed that the rate coefficients of reactions of C$_3$H$^+$ with HD and D$_2$  are different from those with H$_2$. We also report in Table \ref{table:reac_c3hplus} the corresponding rate coefficients used in our network. Let us point out that these values are still subject to some uncertainties as discussed in \cite{SavicGerlich2005} and \cite{Savic2005}.

As in our previous studies of isotopic fractionation \citep{Roueff2015}, we do not solve the full para/ortho chemistry of hydrogen compounds but rather introduce the ortho-to-para ratio of H$_2$, {\small OPR}, as a free parameter. This factor impacts the reverse reactions of the deuteron exchanges of HD with H$_3^+$, H$_2$D$^+$, and D$_2$H$^+$ and the reaction N$^+$ + H$_2$ \citep{Dislaire2012}, as shown in Table~\ref{table:reac_oh2}, where \mbox{$n$($o$-H$_2$) = $\frac{{\small OPR}}{1 + {\small OPR}}  n({\rm{H_2}}$)}. The exponential terms, $\gamma$, were obtained by subtracting  the energy of the $J=1$ level of $o$-H$_2$, 170.5\,K, from the endothermicity of the reaction with reactants and products in their ground states. We did not introduce the similar reactions CH$_{3-n}$D$_n^+$ ($n=0,1,2,3$) + $o$-H$_2$, C$_2$H$_{2-n}$D$_n^+$ ($n=0,1,2$) + $o$-H$_2$, and C$_3$D$^+$ + $o$-H$_2$ since the corresponding endothermicities are much larger \citep{Roueff2013,Nyman2019} and the reactions do not take place in the present low temperature, $\sim$\,10\,K, conditions.

\begin{table}
\small
\caption{H/D ratios of C-containing molecules observed in L483 and steady-state values calculated with our gas-phase chemical model\,$^a$.}
\label{table:model}
\centering
\begin{tabular}{{lrrrr}}
\hline \hline
\multicolumn{1}{|l}{Molecular ratio} &  \multicolumn{1}{c}{L483} & \multicolumn{1}{c}{Model\,1} & \multicolumn{1}{c}{Model\,2} & \multicolumn{1}{c|}{Model\,3} \\
\multicolumn{1}{|l}{OPR H$_2$ ......................}   & & \multicolumn{1}{c}{10$^{-3}$}   & \multicolumn{1}{c}{10$^{-4}$}   & \multicolumn{1}{c|}{10$^{-4}$}   \\
\multicolumn{1}{|l}{$\zeta$ (10$^{-17}$ s$^{-1}$)\,$^d$ ..............} & & \multicolumn{1}{c}{1.3}   & \multicolumn{1}{c}{1.3}   & \multicolumn{1}{c|}{0.5} \\
\hline
 CH$_3$CCH / CH$_2$DCCH   &   5.2\,$^c$ & 10.1 &  7.6   &  6.3 \\
CH$_3$CCH / CH$_3$CCD      &  15.7\,$^c$ & 27.3 & 20.4 & 17.1 \\
CH$_3$CCH/ CHD$_2$CCH    &  33.8\,$^c$ & 91.1 & 56.0 &  38.0 \\
CH$_3$CCH / CH$_2$DCCD   &  42.1\,$^c$ & 131.5 &   78.4   &  54.5 \\
\hline
$c$-C$_3$H / $c$-C$_3$D                &  22.7\,$^d$ & 18.2 & 15.4 &  12.8 \\ 
$c$-C$_3$H$_2$ / $c$-C$_3$HD      &    9.8\,$^d$ & 13.4 & 13.1 &  11.1 \\ 
$c$-C$_3$H$_2$ / $c$-C$_3$D$_2$ &   103\,$^d$ & 173 & 130   &  95.5 \\ 
$l$-C$_3$H$_2$ / $l$-C$_3$HD        &     13.2\,$^d$  & 7.0 & 6.5    &  5.4 \\ 
C$_4$H / C$_4$D                      &     52.6\,$^d$  & 13.0 & 9.7  &  8.9 \\ 
HC$_3$N / DC$_3$N                 &     35.7\,$^d$  & 14.1 & 10.8 & 9.0 \\ 
CH$_3$CN / CH$_2$DCN         &    7.6\,$^d$  & 18.2 & 15.4 & 13.8 \\ 
\hline
\end{tabular}
\tablenotea{$^a$\,See text for the assumed physical conditions. We adopted the elemental abundance ratios \mbox{O/H = $8 \times 10^{-6}$}, \mbox{D/H = $1.5 \times 10^{-5}$}, \mbox{C/O = 0.5}, and \mbox{C/N = 4}. $^b$\,Cosmic-ray ionization rate of H$_2$. $^c$\,Observed values in this work (see Table~\ref{table:column_densities}). $^d$\,Observed values from \cite{Agundez2019}.}
\end{table}

Table \ref{table:model} displays the deuterium ratios of various carbon-containing molecules observed in L483, including present methyl acetylene detections, and three steady-state model results corresponding to the physical conditions, $n$(H$_2$) = $3 \times 10^4$\,cm$^{-3}$ and $T$ = 10\,K, prevailing in L483 \citep{Agundez2019}. Model\,1 represents a fiducial model with {\small OPR} = 10$^{-3}$ and $\zeta = 1.3 \times 10^{-17} s^{-1}$, which corresponds to typical values for dark clouds. We notice that this choice overestimates  the H/D ratio of methyl acetylene by about a factor of two, a feature that we already noticed in our previous study of TMC-1 \citep{Cabezas2021}. Decreasing the ortho-to-para ratio of H$_2$ by a factor of ten, as done in model\,2, allows to decrease reasonably the H/D ratios of the various C-containing molecules. This trend is further strengthened in model\,3, where the cosmic-ray ionization rate is reduced down to $5 \times 10^{-18}$\,s$^{-1}$, which results in H/D ratios for singly and doubly deuterated methyl acetylene close to the observed values.

While a more detailed study is highly desirable, we can make some comments. The abundance ratio between the two singly deuterated forms of methyl acetylene, CH$_2$DCCH/CH$_3$CCD, is close to three according to the observations and this fact is reproduced by the models. This feature results directly from the statistical considerations for the insertion of deuterium in the CH$_3$ group and in the terminal CCH group. For example, in the dissociative recombination of C$_3$H$_6$D$^+$ and C$_3$H$_5$D$^+$, which are the principal formation routes of deuterated methyl acetylene, we explicitly account for this statistical factor in the reaction rate coefficients. The doubly deuterated forms, on the other hand, are assumed to be produced in equal amounts from the dissociative recombinations of C$_3$H$_5$D$_2^+$ and C$_3$H$_4$D$_2^+$, following statistical arguments as well. The difference between the abundances of CH$_2$DCCD and CHD$_2$CCH results then from the destruction rates, for which we assume that the possible ejection of H or D occurs from the terminal CCH or CCD group. As an example, CH$_2$DCCD + H$_3^+$  $\rightarrow$ C$_3$H$_2$D$^+$ + H$_2$ + HD whereas CHD$_2$CCH + H$_3^+$ $\rightarrow$ C$_3$HD$_2$$^+$ + H$_2$  + H$_2$. Then, CHD$_2$CCH is more easily recycled in a doubly deuterated form and its abundance is somewhat larger than that of CH$_2$DCCD, as derived from the observations. A low ortho-to-para ratio of H$_2$ favors significantly the deuterium enhancement of methyl acetylene, which reflects the role of deuteron transfer in ion-molecule reactions. We note that the range 10$^{-3}$ - 10$^{-4}$ is within the {\small OPR} H$_2$ values derived for typical conditions of dark clouds \citep{Lebourlot1991,Furuya2015}. In a similar way, lowering the cosmic-ray ionization rate leads to a decrease in the ionization fraction, which enhances the abundance of molecular ions and decrease the H/D ratio of the various molecular species.

We plan to further investigate the dependence of deuterium fractionation on various parameters in a dedicated study but these first results show that the gas-phase ion-molecule schema can explain the formation of methyl acetylene and its singly and doubly deuterated substitutes. Moreover, if deuteration occurs on grain surfaces and deuterated ratios are governed by the D/H ratio in the gas phase, then it would hold that CH$_2$DCCD/CH$_3$CCD = 3 $\times$ CH$_3$CCD/CH$_3$CCH. That is, CH$_2$DCCD/CH$_3$CCD should be 0.19, while the observed value is 0.37, meaning that doubly deuterated species are twice more abundant than expected if deuteration occurs on grain surfaces following a statistical scheme. This fact further strengthens the gas-phase origin of deuteration for methyl acetylene. It is interesting to note that our models also predict the deuteration of allene (CH$_2$CCH$_2$) at similar levels than methyl acetylene. However, the dipole moment of the fully hydrogenated form of allene is zero and that resulting from deuterium inclusion is probably too small to permit the detection of any deuterium isotopologue.

\section{Conclusions}

We have presented the first detection in space of the two doubly deuterated isotopologues of methyl acetylene, CHD$_2$CCH and CH$_2$DCCD, toward the dark cloud core L483. The frequencies predicted for these two species from laboratory data have significant errors when compared with those derived from astronomical observations. We have therefore derived new spectroscopic parameters for the two deuterium isotopologues which have the accuracy needed to observe these species in cold interstellar clouds showing narrow lines. We derive abundance ratios of CH$_3$CCH/CHD$_2$CCH = $34 \pm 10$ and CH$_3$CCH/CH$_2$DCCD = $42 \pm 13$. Doubly deuterated methyl acetylene is found to be only a few times less abundant than the singly deuterated forms. We have constructed a gas-phase chemical model including multiply deuterated molecules, which is able to reproduce the abundance ratios of the different deuterated isotopologues with respect to CH$_3$CCH observed in L483. In particular, the fact that CHD$_2$CCH is slightly more abundant than CH$_2$DCCD is well accounted for by the chemical model. This favors a scenario in which deuteration of methyl acetylene occurs in the gas phase, rather than on the surface of dust grains.

\begin{acknowledgements}

We acknowledge funding support from Spanish MICIU through grants AYA2016-75066-C2-1-P, PID2019-106110GB-I00, and PID2019-107115GB-C21 and from the European Research Council (ERC Grant 610256: NANOCOSMOS). M.A. also acknowledges funding support from the Ram\'on y Cajal programme of Spanish MICIU (grant RyC-2014-16277). C.C. thanks Prof. Cristina Puzzarini (Universit\'a di Bologna) for her advices and comments about the quantum chemical calculations.

\end{acknowledgements}

\end{document}